# Laser-induced ultrafast electron emission from a field emission tip

B Barwick[1], C Corder, J Strohaber, N Chandler-Smith, C Uiterwaal and H Batelaan

Department of Physics and Astronomy, University of Nebraska—Lincoln,
116 Brace Laboratory, PO Box 880111, Lincoln, Nebraska 68588-0111, USA
E-mail: brettbarwick@gmail.com

**Abstract.** We show that a field emission tip electron source that is triggered with a femtosecond laser pulse can generate electron pulses shorter than the laser pulse duration (~100 fs). The emission process is sensitive to a power law of the laser intensity, which supports an emission mechanism based on multiphoton absorption followed by over-the-barrier emission. Observed continuous transitions between power laws of different orders are indicative of field emission processes. We show that the source can also be operated so that thermionic emission processes become significant. Understanding these different emission processes is relevant for the production of sub-cycle electron pulses.

---

[1] Author to whom any correspondence should be addressed.

The temporal resolution of ultrafast electron diffraction (UED) [1, 2], ultrafast electron microscopy (UEM) [2-4] and ultrafast electron crystallography (UEC) [2] is limited to the duration of the electron pulse. The ultrafast electron source most commonly used in these applications is based on electron emission induced by focusing an amplified femtosecond laser pulse onto a surface [1, 5]. Due to the high particle density per pulse, space-charge broadens the pulse duration to ~500 fs [5]. An interesting electron source implementing a Ti:sapphire oscillator has been used to generate 27-fs electron pulses [6] using impulsively excited surface plasmons. However, this method has a large kinetic energy spread ($\Delta E \sim 100\,\text{eV}$) in the emitted electrons, which would cause the pulse to expand temporally as it propagates. An electron gun has been proposed that could produce sub-fs electron pulses [7]. This ~10 keV source would have an initial energy spread of $\Delta E \sim 1\,\text{eV}$, but this spread would be reduced by injecting the electrons into an RF cavity to compensate for the different velocities, allowing the electron packet to arrive at a target in a sub-fs time window. A promising and experimentally realized source relies on the combination of a field emission tip with a low power femtosecond oscillator [8, 9]. The low laser powers required allow the production of few electrons per pulse with high repetition rates. This gives useful average electron count rates and overcomes space-charge broadening. Field emission tip sources also have small energy spreads ($\Delta E < 1\,\text{eV}$) [10, 11], so their temporal expansion is suppressed. The electron pulses were experimentally shown to be shorter than 100 fs [9]. Assuming that the emission process from the nanometer tip is due to optical field emission, the electrons bunches were claimed to be sub-cycle [9].

In this paper we focus on the nature of the emission process of electrons from a nanometer tip due to femtosecond laser pulses. Our pump-probe results support the claim that the electron pulses are shorter than 100 fs in agreement with Hommelhoff *et al.* [9]. We investigate two possible mechanisms that could describe the emission process. The first is based on the instantaneous laser electric field lowering the potential barrier, thus allowing electrons to tunnel out of the tip (optical field emission). The second mechanism is multi-photon over-the-barrier emission [12]. An analysis of our experimental data shows characteristics of both mechanisms. This is in accordance with the value of our Keldysh parameter [6, 13, 14]. The nature of the emission process is important for speculation on the electron pulse duration. In Refs. [8] and [9] the dominant process is identified as optical field emission, which leads to the prediction that this electron source could produce sub-cycle electron pulses [9]. However, our results indicate a competing process that can be dominant, which stimulates a debate on the temporal characteristics and operating parameters of this source. Only direct experimental evidence, such as diffraction in time experiments [15-17], would unambiguously support the claim that the electron emission is sub-cycle.

To help determine the nature of the emission process we study the two mechanisms (Fig. 1). The optical field emission process is electron-tunneling through a barrier $V$ that has been lowered by $F_{\text{tot}}$, the sum of a DC and laser field. Tunneling is most likely for electrons close to the Fermi level, $E_\text{F}$. Multiphoton absorption can lead to over-the-barrier emission. Upon absorption of four or more photons the gained electron energy exceeds the work function $\phi$ and direct emission can occur. An applied DC field reduces the workfunction to $\phi_{\text{eff}}$ (Schottky effect [18]) thus lowering the number of photons required for over-the-barrier emission. We also consider the possibility of photon absorption followed by tunneling. These models do not include any band structure [12], collision dynamics in the tip [19], or dynamic polarizability in the tip [20-24], and cannot be expected to describe the detailed dynamics of the emission process. However, our simple model agrees well with experiment. We now turn our attention to a more detailed description of the model.

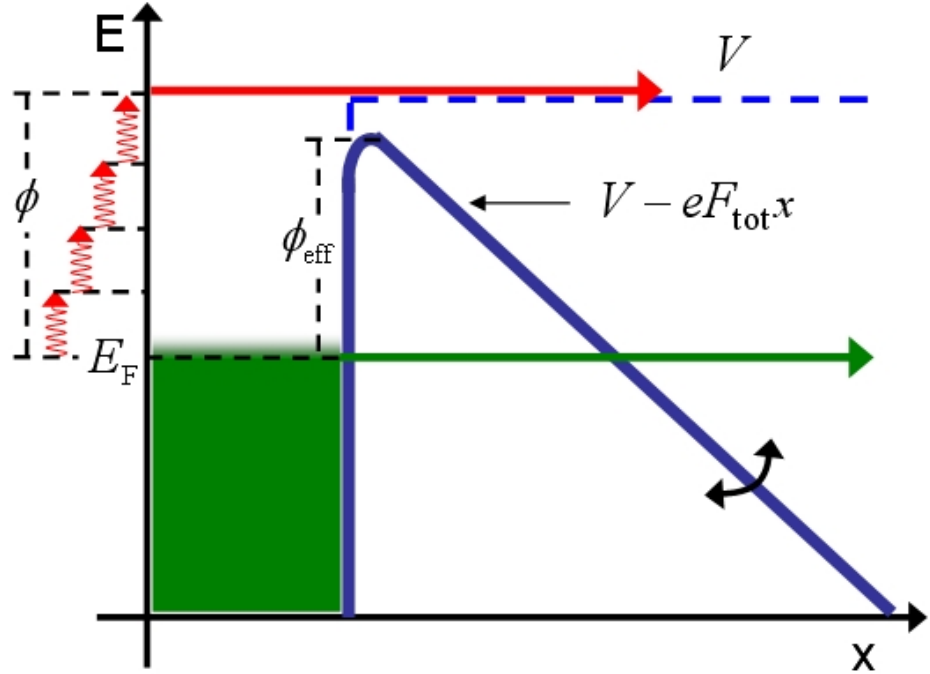

**Figure 1.** Electron emission mechanisms. Optical field emission is depicted with the green arrow and multiphoton absorption followed by over-the-barrier emission is depicted with the red arrow. If an electron at the Fermi energy, $E_F$, absorbs multiple photons it can gain enough energy to exceed the potential well. The height of the effective potential barrier is determined by the work function of the metal $\phi$ and the applied DC field (Schottky effect [18]). Photoexcitation may precede tunneling (for a detailed description see text).

For $n$-photon absorption the electron emission rate $J_{abs,n}$ is proportional to the $(2n)$-th power of the laser field ($\theta$ is the angle between the laser field polarization and the tip axis),

$$J_{abs,n}(F_{laser}) \propto (F_{laser} \cos\theta)^{2n}, \qquad (1)$$

where the energy of the excited electron must exceed the effective potential barrier, $n\hbar\omega > \phi_{eff}$. The Schottky effect is given by $\phi_{eff} = \phi - e\left[eF_{DC}/(4\pi\varepsilon_0)\right]^{1/2}$ [18]. Above-threshold photoemission ($n \geq 5$) could contribute to the total electron emission signal [25].

The polarization ($\theta$) dependence of the electron emission can be attributed to an increased probability of photon absorption when the electric field is perpendicular to a surface (parallel to the tip axis) [26]. Other groups have attributed the polarization dependence to the motion of the conduction electrons in the metal tip [21-24]. This *lightning effect* treatment causes an enhancement of the field near the tip because the optical field could make conduction electrons bunch at the tip apex. There is considerable disagreement in both theoretical and experimental papers [21-24] to the amount of enhancement, but, fortunately, the lightning effect does not affect the power law behavior and is therefore not explicitly given in Eq. (1).

In our pump-probe experiments, the temporal dependence of the laser field is given by

$$F_{\text{laser}}(t) = F_{01}\exp(-at^2)\cos(bt^2 + \omega t) + $$
$$+ F_{02}\exp\left(-a(t+x_0/c)^2\right)\cos\left(b(t+x_0/c)^2 + \omega(t+x_0/c)\right), \quad (2)$$

where $a$ and $b$ are the complex beam parameters [27]. This laser field is the superposition of two pulses, separated in time by a relative delay of $x_0/c$ which we introduce using an autocorrelator. We determine the magnitudes $F_{0i}$ of each of the two pulses separately by measuring their average laser power $P_{\text{avg}}$:

$$F_{0i} = \sqrt{\frac{16 P_{\text{avg},i}}{f_{\text{rep}} d^2 t_{\text{laser}} \varepsilon_0 c (\pi/\ln 2)^{3/2}}} \approx 25 \sqrt{\frac{P_{\text{avg},i}}{f_{\text{rep}} d^2 t_{\text{laser}}}}, \quad (3)$$

in which $f_{\text{rep}}$ is the repetition rate of the laser, $d$ the full-width-half-maximum of the focal spot, and $t_{\text{laser}}$ the laser pulse duration. We find the emission current as a function of the autocorrelator delay $x_0/c$ by numerical integration over time. Equation (1) directly gives the electron emission as a function of polarization and laser power.

If the laser field causes electrons to tunnel, the current from the tip, $J_{\text{field}}$, is given by the Fowler-Nordheim equation,

$$J_{\text{field}}(F_{\text{tot}}) \propto \frac{a_n F_{\text{tot}}^2}{(\phi - n\hbar\omega)} \exp\left(-C_0 (\phi - n\hbar\omega)^{3/2} / F_{\text{tot}}\right), \quad (4)$$

for a constant $C_0$ [28], and the excited state populations are assumed to be proportional to a power of the laser intensity, $a_n \propto I^n$. The total electric field is the sum of the static and laser fields: $F_{\text{tot}} = F_{\text{DC}} + F_{\text{laser}} \ell \cos\theta$. The factor $\ell$ accounts for the field enhancement due to the lightning effect. The static electric field is given by $F_{\text{DC}} = V/kr$, where $V$ is the voltage placed on the tip, $r$ is the tip radius, and $k$ (here equal to 5) accounts for details in the tip geometry [29]. To describe tunneling preceded by absorption of $n$ photons we lower the workfunction to $\phi - n\hbar\omega$. We find the emission current by numerical integration over time.

Such time averaging is only appropriate if the tunneling time of the electron from the metal tip ([14]) is shorter than the optical period of the laser light. In the field of an electromagnetic wave, the critical parameter identifying this regime is the Keldysh parameter $\gamma$. For $\gamma \gg 1$ multiphoton absorption dominates, while tunneling dominates for $\gamma \ll 1$ [13, 14]. For a metal surface the Keldysh parameter is given by $\gamma = \omega(2m\phi)^{1/2}/(eF_{\text{laser}})$ [6]. For a field emission tip the magnitudes of the DC field and laser field are of the same order and the emission process is strongly dependent on both fields. The Keldysh parameter should thus depend not only on $F_{\text{laser}}$ but also on $F_{\text{DC}}$. This dependence is present when the workfunction is replaced with the effective workfunction. Note that the Keldysh parameter is only meaningful when the photon energy is less than the workfunction.

We also consider the possibility of laser induced thermionic emission. It has been suggested that thermionic emission is most efficient when the laser polarization is perpendicular to the tip axis [30]. On the other hand, in the investigated experimental regime thermionic emission is thought to be negligible [20, 28, 30].

A schematic of our experimental setup appears in Fig. 2. Laser pulses from a Ti:sapphire femtosecond oscillator (Spectra Physics Tsunami) are first compressed using a single-prism pulse compressor [31]. A subsequent variable attenuator controls the delivered laser power. We use a Mach-Zehnder type autocorrelator to split the laser pulse into a pump and probe pulse and to provide a variable time delay between them. We let the recombined laser beam emanating from the autocorrelator pass through a half-wave plate to adjust the overall laser polarization. A frequency resolved optical gate (FROG) is used to measure the pulse characteristics just before the pulse enters the vacuum system. We measure a laser pulse width of 32 fs, with a time-bandwidth product of 0.5. The fused silica vacuum entrance window is 3 mm thick. To focus the laser beam onto the field emission tip we use a 90-degree off-axis gold-coated parabolic mirror placed in the vacuum system (parent focal length = 12.7 mm, P/N A8037-176 Janos Technology). We have connected the field emission tip to an *XYZ* translation stage through a flexible bellows to allow for optimization of the electron emission. The tungsten tip is etched with a lamella drop-off method [32].

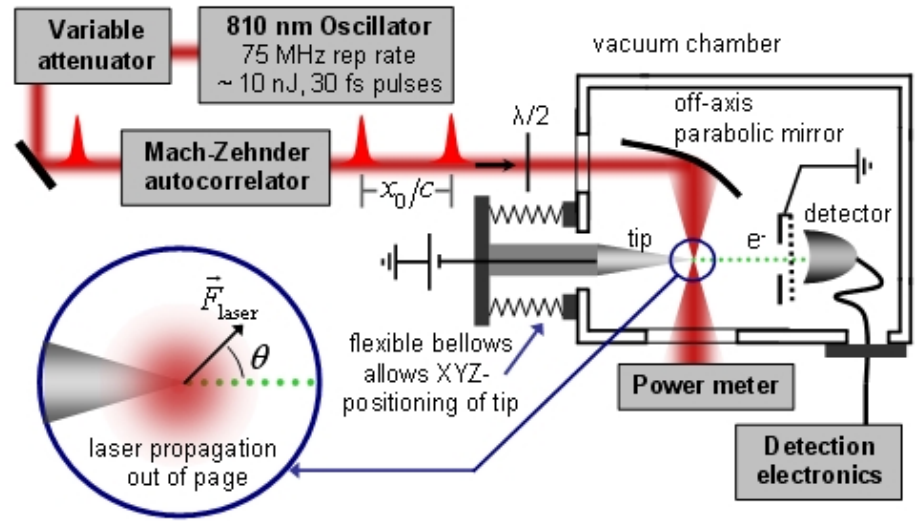

**Figure 2.** Experimental setup. A femtosecond laser oscillator produces radiation pulses at a rate of 75 MHz. An autocorrelator provides time adjustable pump and probe pulses. The laser pulses are focused on a field emission tip to extract electron pulses. (For a more detailed description see text.)

The experiment is contained in an aluminum vacuum chamber that is evacuated with a turbomolecular pump and is operated at a pressure of $\sim 10^{-8}$ Torr. To estimate the tip radius the Fowler-Nordheim equation is fit [29] to the voltage-dependent electron emission yield, giving a value of ~40 nm. A metal plate with a 5-mm pinhole placed at 1 cm from the tip defines the ground potential. A channeltron (Sjuts KBL 520) is used to detect the electrons. Once the laser pulses have left the vacuum system through a second optical window we measure the average beam power with a power meter. The electron pulse detection signals are sent through a constant fraction discriminating amplifier, and then fed into a multichannel scaling board. This board records electron emission autocorrelation spectra, which provides information on the femtosecond scale. The laser intensity autocorrelation trace (Fig. 3a) is measured simultaneously with the electron emission autocorrelation spectrum (Fig. 3b). In this way the laser intensity can be correlated to the electron emission. The oscillator delivers pulses at a repetition rate of 75 MHz, with a maximum average output power of ~500 mW corresponding to a pulse energy of ~10 nJ. This is enough to damage the tip so a variable attenuator is used. To avoid detector damage we limit the electron count rate to $10^6$/s (reached at ~25 mW depending on the applied DC voltage). We estimate the laser focus to have a full width at half maximum of ~ 4 μm.

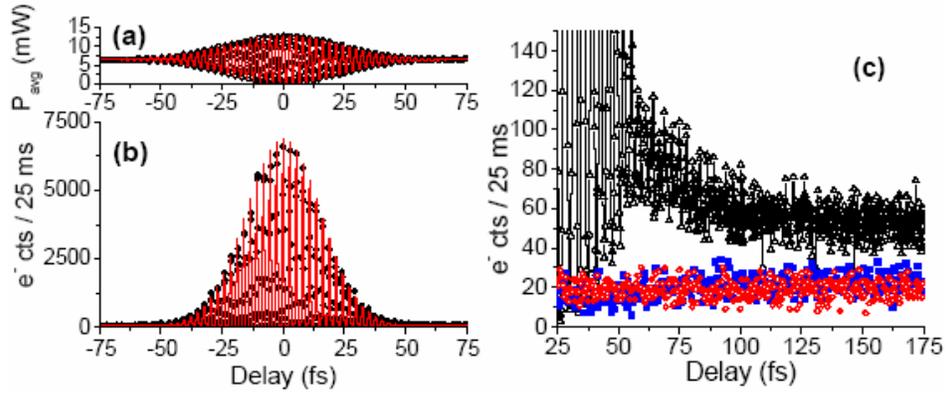

**Figure 3**. Autocorrelation traces and pump-probe measurements. Panel (a) depicts the average laser power as a function of delay between the pump and probe pulse. Panel (b) shows the corresponding electron counts. The red line on the electron data is a fit proportional to the 4$^{th}$ power of the intensity (laser pulse 50 fs, $V_{DC} = -50$ V, corresponding DC field 0.25 GV/m). Panel (c) shows an extension of the right wing of the autocorrelation trace. The red circles and the blue squares are the electron counts when the pump and probe pulse are blocked, respectively. The black triangles were measured with both pulses present. The sum of the pump and probe signals is approximately equal to that with both pulses present for large delays. This indicates that the laser pulses produce electrons independently.

When the two pulses emanating from the autocorrelator are delayed to the extent that they do not overlap, they act as a pump-probe pair. The first pulse influences the tip and the second pulse probes if the tip "remembers" the first pulse. Once the electron emission from the two laser pulses becomes additive, there is no memory of the first pulse anymore, so the electron emission process should be at least as fast as the delay for which this happens (Fig. 3c). The temporal resolution of this pump-probe experiment is limited to the duration of the laser pulse, because the two pulses are coherent with each other. As the delay between them is reduced below their temporal width, the two laser pulses start to interfere, creating an intensity modulation which is seen in the autocorrelation trace (Fig. 3a).

At delays greater than 100 femtosecond the sum of the electron signals with each laser pulse separately (blue and red), nearly equals the electron signal with both pulses present (black). As discussed above, this additive behavior indicates that the electron emission process is faster than 100 fs. In a very recent similar study [9] this additive behavior is also reported at 100 fs.

A duration of 100 fs is compatible with the mechanisms we consider, except for thermionic emission. For a field emission model one expects that the electron emission process is of sub-cycle duration [9]. The time for multiphoton over-the-barrier emission for a tip is unknown to the authors and will depend on the detailed electron dynamics.

To help identify the physical process(es), the laser intensity dependence of the emission yield is measured. The experimental data are shown in Fig. 4a on a log-log scale. Our data can not be fit to a pure optical tunneling model from the Fermi level; this contrasts with the interpretation given in reference [8]. To investigate the n-photon absorption model in detail (Eq. (1)), we fit our data to $\sum_{n=0,5} c_n I^n$. The data labeled with "n=4" in Fig. 4a is dominated by the fourth order power law, consistent with a very recent observation [33]. The data labeled with "n=3" in Fig. 4a has significant contributions from the second, third and fourth order (Fig. 4b). Varying the DC-field shows that contributions from more than one order is typical. As a result, a single power fit to the electron yield shows that integer powers are uncommon (Fig. 4c). Optical field emission described by Eq. (4) including excited state population can also appear as a straight line (see dashed line (FN) in Fig. 4a). All these observations render an identification of the specific process difficult.

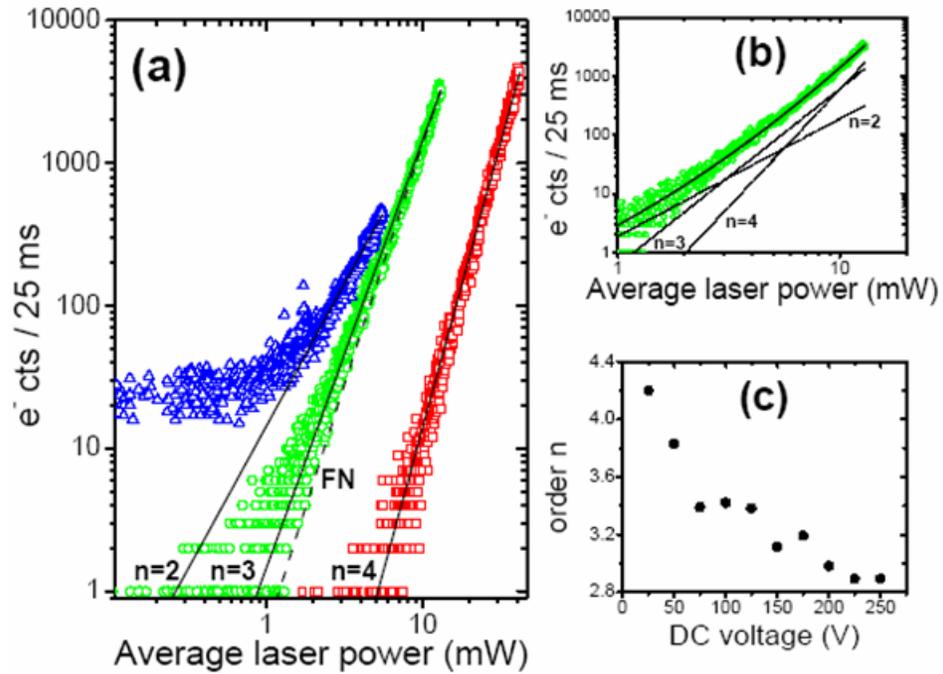

**Figure 4.** Multiphoton order of the electron emission. Electron counts versus the time-averaged laser power taken at different DC tip voltages are shown in panel (a) on a log-log scale. The blue triangle data points were recorded with a DC voltage of −450 V (field ~2.25 GV/m), the green circles with −300 V (~1.5 GV/m) and the red squares with −50 V (~0.25 GV/m). The three straight black lines are $\propto I^n$ for n=2,3,4, from left to right respectively. The data points corresponding to −450 $V_{DC}$ (blue triangles) approach a constant count rate at low laser power due to DC Fowler-Nordheim emission. Given the FWHM pulse duration of 50 fs, a 40 mW average laser power corresponds to a peak laser field of 0.6 GV/m. Panel (b) shows how the electron emission can be broken down in the contributions of the separate multiphoton orders (for the "n=3" data in panel (a)). Panel (c) shows the result of a fit to a single power law for various DC fields. An integer power is the exception.

For our experiment, the observed $I^4$ dependence may be due to two different mechanisms. The first mechanism is absorption of four 810-nm photons with a combined energy of ~ 6.0 eV, which equals or exceeds the value of all known work functions for tungsten (between 4.3 and 6.0 eV) [34]). A second explanation could be that three photons are sufficient to cause over-the-barrier emission, with a fourth photon being absorbed in above-threshold photoemission [12, 25].

Our estimated Keldysh parameter is consistent with our observations. For an average laser power of 40 mW (the highest power in figure 4(a)) and an FWHM pulse duration of 50 fs, the optical field is 0.6 GV/m. This field is nearly equal to that used in reference [6] and [8]. For the range of DC fields used, the Keldysh parameter is between 3 and 4. The photon energy is less than the effective workfunction which justifies the Keldysh approach.

Finally, a polarization dependent measurement can reveal the presence of laser induced thermionic emission. As mentioned above, thermionic emission is thought to be most efficient when the laser polarization is perpendicular to the tip axis [30]. A half wave plate is used to rotate the laser polarization (Fig. 5). In Fig. 5a the electron emission is negligible for polarization perpendicular to the tip. This is achieved by carefully aligning the apex of the tip to coincide with the center of the laser focus, which is the chosen configuration for all the data taken above. However, when the tip is moved slightly deeper into the focal region a secondary peak appears for perpendicular polarization (Fig. 5b).

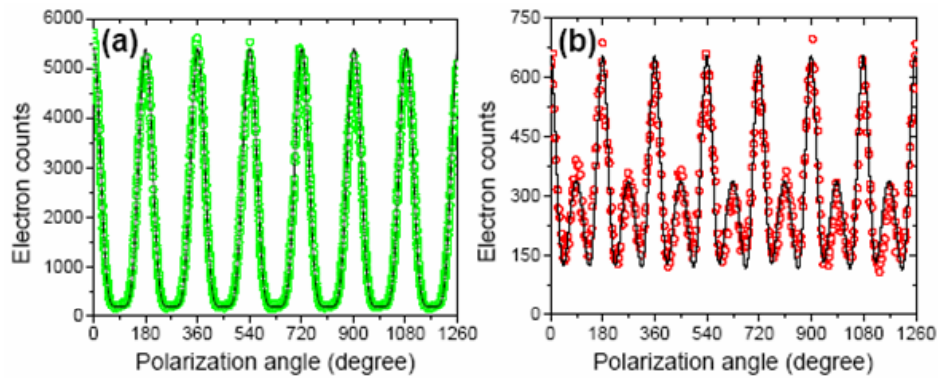

**Figure 5.** Polarization dependent electron yield. Panel (a) shows the electron signal as a function of laser polarization angle for careful alignment. A maximum yield is observed when the laser polarization is parallel to the tip, and the minima correspond to laser polarizations perpendicular to the tip. This data is taken for the same experimental condition as the "n=3" data in Fig. 4a. Panel (b) shows a typical measurement with a less careful alignment which leads to a secondary peak for perpendicular polarization. The black lines are guides to the eye.

In conclusion, the additive nature of the pump-probe experiment is strong evidence that the electron emission process is prompt to the ~100 fs level in agreement with Hommelhoff *et al.* [9]. Sub-100-fs resolution is promising for ultrafast electron microscopy and diffraction. One of the factors limiting the currently achieved resolution to about 1 ps is space charge broadening [5], which is absent for this electron source. Electrons from a 40-nm tip source can be focused for ultrafast microscopy and collimated for ultrafast diffraction.

Our experiment shows characteristics of multiple emission processes. Experimental parameters can be chosen to favor a particular process. This opens perspectives to optimize our source to attain the shortest possible electron pulses duration. Diffraction-in-time methods may offer the possibility to monitor the optimization process down to the attosecond domain.
**Acknowledgements**

This material is based upon study supported by the National Science Foundation under Grant No. 0354940 and Grant No. 0355235.

**References**

[1] Dantus M, Kim S B, Williamson J C and Zewail A H 1994 *J. Phys. Chem* **98** 2782
[2] Zewail A H 2006 *Annu. Rev. Phys. Chem.* **57** 65
[3] Lobastov V A, Srinivasan R and Zewail A H 2005 *PNAS* **102** 7069
[4] King W E, Campbell G H, Frank A, Reed B, Schmerge J F, Siwick B J, Stuart B C and Weber P M 2005 *J. Appl. Phys.* **97** 111101
[5] Siwick B J, Dwyer J R, Jordan R E and Miller R J D 2003 *Science* **302** 1382
[6] Irvine S E, Dechant A and Elezzabi A Y 2004 *Phys. Rev. Lett.* **93** 184801
[7] Fill E, Veisz L, Apolonski A and Krausz F 2006 *New J. Phys.* **8** 272
[8] Hommelhoff P, Sortais Y, Aghajani-Talesh A and Kasevich M 2006 *Phys. Rev. Lett.* **96** 077401
[9] Hommelhoff P, Kealhofer C and Kasevich M A 2006 *Phys. Rev. Lett.* **97** 247402
[10] Kiesel H, Renz A and Hasselbach F 2002 *Nature* **418** 392
[11] Gadzuk J W and Plummer E W 1973 *Rev. Mod. Phys.* **45** 487
[12] Bisio F, Hývlt M, Franta J, Petek H and Kirschner J 2006 *Phys. Rev. Lett.* **96** 087601
[13] Keldysh L V 1965 *Sov. Phys. JETP* **20** 1307
[14] Rzazewski K and Roso-Franco L 1993 *Laser Phys.* **3** 310
[15] Lindner F, Schätzel M G, Walther H, Baltuška A, Goulielmakis E, Krausz F, Milošević D B, Bauer D, Becker W and Paulus G G 2005 *Phys. Rev. Lett.* **95** 040401
[16] Colombe Y, Mercier B, Perrin H and Lorent V 2005 *Phys. Rev. A.* **72** 061601
[17] Szriftgiser P, Guéry-Odelin D, Arndt M and Dalibard J 1996 *Phys. Rev. Lett.* **77** 4
[18] Coulombe S and Meunier J-L 1997 *J. Phys. D: Appl. Phys.* **30** 776
[19] Lea C and Gomer R 1970 *Phys. Rev. Lett.* **25** 804
[20] Vurpillot F, Gault B, Vella A, Bouet M and Deconihout B 2006 *Appl. Phys. Lett.* **88** 094105
[21] Gault B, Vurpillot F, Bostel A, Menand A and Deconihout B 2005 *Appl. Phys. Lett.* **86** 094101
[22] Novotny L, Bian R X and Xie X S 1997 *Phys. Rev. Lett.* **79** 645
[23] Martin Y C, Hamann H F and Wickramasinghe H K 2001 *J. Appl. Phys.* **89** 5774
[24] Bouhelier A 2006 *Microsc. Res. Tech.* **69** 563
[25] Banfi F, Giannetti C, Ferrini G, Galimberti G, Pagliara S, Fausti D and Parmigiani F 2005 *Phys. Rev. Lett.* **94** 037601
[26] Venus D and Lee M J G 1983 *Surface Science* **125** 452
[27] Siegman A E 1986 *Lasers* (Sausalito: University Science Books)
[28] Binh V T, Garcia N and Purcell S T 1996 *Advances in Imaging and Electron Physics,* ed P W Hawkes (San Diego: Academic Press) pp 63-153
[29] Gomer R 1961 *Field Emission and Field Ionization* vol 9 (Cambridge: Harvard University Press)
[30] Hadley K W, Donders P J and Lee M J G 1985 *J. Appl. Phys.* **57** 2617
[31] Akturk S, Gu X, Kimmel M and Trebino R 2006 *Opt. Express* **14** 10101

[32] Müller A-D, Müller F, Hietschold M, Demming F, Jersch J and Dickmann K 1999 *Rev. Sci. Instr.* **70** 3970
[33] Ropers C, Solli D R, Schulz C P, Lienau C and Elsaesser T 2007 *Phys. Rev. Lett.* **98** 043907
[34] Müller E W 1955 *J. Appl. Phys.* **26** 732